\documentclass[12pt,prd,tightenlines,nofootinbib,showpacs,showkeys]{revtex4}
\newcommand{\be}{\begin{equation}}
\newcommand{\ee}{\end{equation}}
\usepackage{bm}
\usepackage{graphics}
\usepackage{rotating}
\begin{document}
\title{\begin{flushright}{\rm\normalsize SSU-HEP-07/8}\end{flushright}
Ground-state triply and doubly heavy baryons\\ in a relativistic three-quark
model}
\author{A. P.
Martynenko\footnote{E-mail:~mart@ssu.samara.ru}}
\affiliation{Samara State University, Pavlov Street 1, Samara 443011,
Russia}

\begin{abstract}
Mass spectra of the ground-state baryons consisting of three or two heavy
($b$ or $c$)
and one light $(u,d,s)$ quarks are calculated in the framework of the
relativistic quark model and the hyperspherical expansion. The predictions
of masses of the triply and doubly heavy baryons are obtained by employing the
perturbation theory for the spin-independent and spin-dependent parts
of the three-quark Hamiltonian.
\end{abstract}



\maketitle

The transition from two-quark bound states to three-quark bound
states opens new problems which refer both to the form of the quark
interaction in the baryon and the structure of three-quark
relativistic wave equation describing this system. In general form
they were studied and solved already by many authors
\cite{CI,KSt,Richard,KKP,KL,QWG,Bali,JLB} (see other references in
Ref.\cite{QWG}). In practice it is important to have an approach
which can allow to obtain simple and reliable estimates for the
different experimental quantities regarding to the baryon
spectroscopy, the production and decay rates. Whereas heavy baryons
with one or two heavy quarks were investigated both theoretically
and experimentally, the triply heavy baryons $\Omega_{Q_1Q_2Q_3}$
containing $b-$ and $c-$ quarks have not studied so much. The
estimate of masses of the lowest-lying $(ccc)$, $(ccb)$, $(bbc)$ and
$(bbb)$ states is presented in Refs.\cite{kuti,bjorken,jia}. Their
production in a $c-$ or $b-$ quark fragmentation is calculated in
Refs.\cite{saleev,gn}. The doubly heavy baryons $(Q_1Q_2q)$
represent a unique part of three-quark systems. Two heavy quarks
compose a localized quark nucleus while the light quark moves around
this color source at a distance of order $(1/m_q)$. This picture
leads to the quark-diquark model for doubly heavy baryons which was
used in Refs.\cite{rqm1,rqm2,rqm3} for the description of the mass
spectrum and decay widths. Moreover, relativistic and bound state
corrections to the mass spectra of mesons and baryons (in the
quark-diquark approximation) and their different decay rates were
also considered in the relativistic quark model
\cite{rqm1,rqm2,rqm3,rqm4,rqm5}. Estimates for the masses of baryons
containing two heavy quarks have been presented by many authors
\cite{NT1,NT2,NT3,GKLO,Roncaglia,KKP} using different QCD inspired
models for the quark interactions. The aim of the present paper is a
twofold one. First, we go beyond the scope of the quark-diquark
approximation in Refs.\cite{rqm1,rqm2} treating the total baryon
Hamiltonian as a sum of two-quark interactions and using the
hyperradial approximation for the ground state triply and doubly
heavy baryons. Secondly, we take into account relativistic and bound
state corrections of order $v^2/c^2$ by the perturbation theory. So,
the purpose of our new investigation consists in the elaboration of
an alternative calculational scheme of the baryon mass spectrum as
compared with the earlier performed investigations in
Refs.\cite{rqm1,rqm2} through the use of the three-quark approach to
the baryon problem formulated in Refs.\cite{NT1,NT2,NT3} with the
Hamiltonian containing the spin-independent and spin-dependent
corrections of order $v^2/c^2$.

It should be noted, that the theoretical results on the mass
spectrum of triply and doubly heavy baryons \cite{KL,QWG,Bali}
remain untapped. In the past several years, the SELEX
Collaboration has reported the first observation of doubly charmed
baryons \cite{SELEX1,SELEX2}. But most recently, BaBar
Collaboration has reported that they have not founded any
evidence of doubly charmed baryons in $e^+e^-$ annihilation \cite{BaBar}.
Nevertheless, we can expect that the mass spectra and decay rates of triply and
doubly heavy baryons will be measured before long. This gives
additional grounds for new theoretical investigations of triply and doubly
heavy baryon properties.

In order to describe the mass spectra of baryons the different
three-quark Hamiltonians are used \cite{rqm1,NT1,Bali,Brambilla}.
The effective Hamiltonian of Refs.\cite{NT1,NT2,NT3} devoted to the
calculation of the mass spectra of doubly heavy baryons in the
three-body approach is a sum of the string potential $V^{conf}$ and
the Coulomb interaction potential $V^C$. A consistent derivation of
the three-quark potential was done in the Wilson-loop approach in
Ref.\cite{Brambilla,BVR} which accounts for corrections of order
$1/m^2$. Accounting for the growth of the strong coupling constant
for the interaction of a light quark with the heavy quark, we have
included in the pure nonrelativistic three-quark Hamiltonian the
vacuum polarization corrections of order $\alpha_s^2$ in the form
\cite{BPSV,KPSS}:
\begin{equation}
H_0=\frac{{\bf p}_1^2}{2m_1}+\frac{{\bf p}_2^2}{2m_2}+ \frac{{\bf
p}_3^2}{2m_3}-\frac{2}{3}\sum_{i<j}\frac{\alpha_s^{ij}} {|{\bf
r}_{ij}|}\left[1+\frac{\alpha_s^{ij}}{4\pi} \left(\tilde a_1+
8\gamma_E\beta_0+8\beta_0\ln (\tilde\mu_{ij}|{\bf r}_{ij}|)
\right)\right]+\sum_{i=1}^3\frac{1}{2}A(r_i+B).
\end{equation}
We take the Hamiltonian (1) as the initial approximation in our study of the
baryon mass spectrum. The operator of quark momenta ${\bf p}_i=
-i\frac{\partial}{\partial{\bf r}_i}$,
${\bf r}_i$ is the position of quark $i$ with respect to a common string-junction
point which coincides approximately with the center-of-mass point (for a more
detailed discussion see Refs.\cite{NT1,NT2,NT3}) and ${\bf r}_{ij}=
{\bf r}_i-{\bf r}_j$ (compare with Fig.1). The parameters of the confinement
part of the potential are the following \cite{rqm1,rqm4,rqm5}:
$A=0.18~GeV^2$, $B=-0.16~GeV$. We take the quark masses $m_b=4.88~GeV$,
$m_c=1.55~GeV$, $m_{u,d}=0.33~GeV$, $m_s=0.5~GeV$ as in our previous \cite{rqm2,rqm3,rqm5}
calculations of different hadron properties on the basis of the relativistic
quark model. For the dependence of the QCD coupling constant $\alpha_s^{ij}=
\alpha_s(\tilde\mu_{ij}^2)$ on the renormalization point $\tilde\mu_{ij}^2$ we use the
two-loop result \cite{CKS}
\begin{equation}
\alpha_s(\tilde\mu_{ij}^2)=\pi\left[\frac{1}{\beta_0 L}-\frac{\beta_1\ln L}
{\beta_0(\beta_0L)^2}\right],~~
\beta_0=\frac{1}{4}\left(11-\frac{2}{3}n_f\right), \beta_1=\frac{1}{16}\left(
102-\frac{38}{3}n_f\right), L=\ln(\tilde\mu_{ij}^2/\Lambda^2),
\end{equation}
where $\Lambda=0.168~GeV$, $\tilde\mu_{ij}=2m_im_j/(m_i+m_j)$ is the
renormalization scale, $\tilde a_1=(31-10n_f/3)/3$
and $n_f$ is the number of flavours. The three-body
Hamiltonian (1) and the Schr\"odinger equation can be reduced
to the two-body form. For this aim, the three-body Jacobi coordinates become
useful
\cite{S1,S2,NT1}:
\begin{equation}
{\mathstrut\bm\rho}_{ij}=\alpha_{ij}({\bf r}_i-{\bf r}_j),~
{\mathstrut\bm\lambda}_{ij}=\beta_{ij}\left(\frac{m_i{\bf
r}_i+m_j{\bf r}_j} {m_i+m_j}-{\bf r}_k\right),
\end{equation}
where the coefficients $\alpha_{ij}$, $\beta_{ij}$ are expressed in terms
of the appropriate reduced masses:
\begin{equation}
\alpha_{ij}=\sqrt{\frac{\mu_{ij}}{\mu}},~\beta_{ij}=\sqrt{\frac{\mu_{ij,k}}
{\mu}},~\mu_{ij}=\frac{m_im_j}{(m_i+m_j)},~\mu_{ij,k}=\frac{(m_i+m_j)m_k}
{(m_i+m_j+m_k)},
\end{equation}
$\mu$ is an arbitrary mass parameter which disappears in final expressions.
Evidently, the coordinate ${\mathstrut\bm\rho}_{ij}$ is proportional to the
distance between quarks $i$ and $j$, and the coordinate
${\mathstrut\bm\lambda}_{ij}$ is proportional to the distance between
the quark $k$ and the center-of-mass of quarks $i,j$. Together with the
center-of-mass coordinate ${\bf R}_{c.m.}$ the Jacobi coordinates determine
completely the position of the system.

In the center-of-mass frame $({\bf R}_{c.m.}=0)$ the operator of the kinetic
energy can be written in the Jacobi coordinates ${\mathstrut\bm\rho},
{\mathstrut\bm\lambda}$ as follows:
\begin{equation}
T_0=-\frac{1}{2\mu}\left(\frac{\partial^2}{\partial{\mathstrut\bm\rho}^2}+
\frac{\partial^2}{\partial{\mathstrut\bm\lambda}^2}\right)=-
\frac{1}{2\mu}\left(\frac{\partial^2}{\partial R^2}+\frac{5}{R}\frac{\partial}
{\partial R}+\frac{K^2(\Omega)}{R^2}\right),
\end{equation}
where $K^2(\Omega)$ is the angular momentum operator, whose eigenfunctions
(hyperspherical harmonics) are determined by the equation \cite{Vilenkin}:
\begin{equation}
K^2(\Omega)Y_K(\Omega)=-K(K+4)Y_K(\Omega).
\end{equation}
Here $\Omega$ designates five angular coordinates and $R$ is the six-dimensional
hyperradius
\begin{equation}
R=\sqrt{{\mathstrut\bm\rho}_{ij}^2+{\mathstrut\bm\lambda}_{ij}^2},~~\rho=R\cos\theta,~~
\lambda=R\sin\theta.
\end{equation}
The baryon wave function $\Psi({\mathstrut\bm\rho},{\mathstrut\bm\lambda})$
can be presented as an expansion over functions $Y_K(\Omega)$:
\begin{equation}
\Psi({\mathstrut\bm\rho},{\mathstrut\bm\lambda})=\sum_K\Psi_K(R)Y_K(\Omega).
\end{equation}

\begin{figure}[htbp]
\centering
\includegraphics{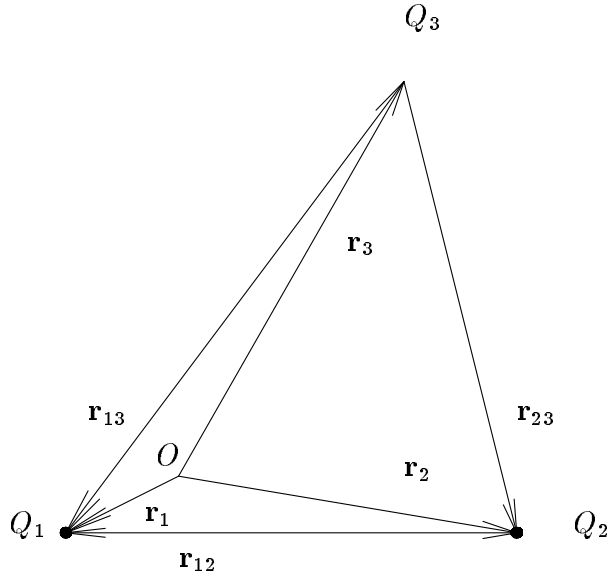}
\caption{The configuration of the three-quark system $(Q_1Q_2Q_3)$.
O is the string-junction point. $Q_1,Q_2$ are the heavy quarks $b$
or $c$. The quark $Q_3$ is treated as a heavy quark $b,c$ in the
triply heavy baryon or a light quark $q$ in the doubly heavy
baryon.}
\end{figure}

Appearing here the radial wave functions $\Psi_K(R)$ satisfy to the system of
differential equations \cite{S1,S2}. However,
in the further study of the ground state triply and doubly heavy baryons we
use the hyperradial approximation in which $K=0$ and the bound state
wave function $\Psi=\Psi(R)$ does not depend on the angular
variables in the six-dimensional space. Using the Jacobi coordinates
the Schr\"odinger equation for the three-quark system can now be
transformed into the following form:
\begin{equation}
\Biggl[-\frac{1}{2\mu}\left(\frac{d^2}{dR^2}+\frac{5}{R}\frac{d}{dR}\right)-
\frac{2}{3}\sum_{i<j}\frac{\alpha_s^{ij}\alpha_{ij}}{|{\mathstrut\bm\rho}_{ij}|}
+\sum_{i<j}\frac{1}{2}\left(A\gamma_{ij}|{\mathstrut\bm\lambda}_{ij}|+B\right)-
\end{equation}
\begin{displaymath}
-\frac{1}{6\pi}\sum_{i<j}\frac{{\alpha_s^{ij}}^2\alpha_{ij}}{|{\mathstrut\bm\rho_{ij}|}}
\left(\tilde a_1+ 8\gamma_E\beta_0+8\beta_0\ln
\frac{(\tilde\mu_{ij}|{\mathstrut\bm\rho}_{ij}|)}{\alpha_{ij}}\right)\Biggr]\Psi(R)=E\Psi(R),
\end{displaymath}
where coefficients $\gamma_{ij}$ are given by
\begin{equation}
\gamma_{ij}=\sqrt{\frac{\mu(m_i+m_j)}{m_k(m_1+m_2+m_3)}}.
\end{equation}
Averaging Eq.(9) over angular variables, we can present the
Schr\"odinger equation for the reduced radial wave function
$\chi(R)=R^{5/2}\Psi(R)$ in the form of a two-body wave equation:
\begin{equation}
\frac{d^2\chi(R)}{dR^2}+2\mu\left[E-V(R)\right]\chi(R)=0,
\end{equation}
\begin{equation}
V(R)=\sum_{k=-2}^{1}a_kR^k+a_2\frac{\ln
R}{R},~a_0=\frac{3}{2}B,~a_1=\frac{16}{15\pi}A\sum_{i<j}\gamma_{ij},~
a_{-2}=\frac{15}{8\mu},
\end{equation}
\begin{displaymath}
a_{-1}=-\frac{32}{9\pi}
\sum_{i<j}\alpha_s^{ij}\alpha_{ij}\left[1+\frac{\alpha_s^{ij}}{\pi}
\left(-\frac{17}{4}+\frac{9}{2}\ln
2+\frac{9}{2}\gamma_E+\frac{9}{2}\ln
\frac{\tilde\mu_{ij}}{\alpha_{ij}}\right)\right],~a_2=-\frac{16}{\pi^2}
\sum_{i<j}{\alpha_s^{ij}}^2\alpha_{ij}.
\end{displaymath}

The Schr\"odinger equation (11) determines the baryon mass
spectrum in the initial approximation. It can be solved
numerically by the use the Mathematica program \cite{schroe}. The
baryon wave function $\chi(x)$ ($x=R\cdot\sqrt{\mu}$) for the state
$(ccc)$ is shown in Fig.2.

\begin{figure}[htbp]
\centering
\includegraphics{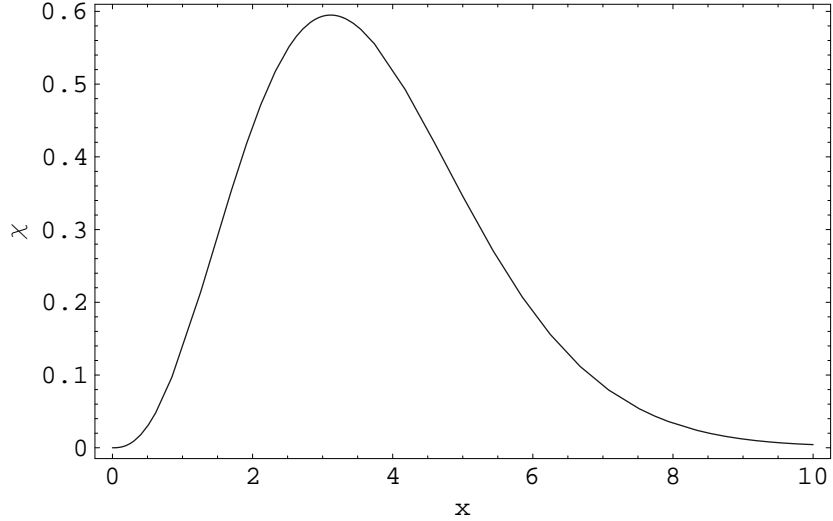}
\caption{The wave function of the baryon $(ccc)$ obtained after numerical
solution of the Schr\"odinger equation (11).}
\end{figure}

The next step in the solution of the spectral problem is related
to the consideration of different corrections to the Hamiltonian
$H_0$. In the present study we suggest that the three-quark potential
has the form of the sum of twin quark the Breit-like interactions (we
neglect orbital motion of the quarks) \cite{rqm2,rqm3,rqm4,rqm5}:
\begin{equation}
V^{SI}({\bf r}_1,{\bf r}_2,{\bf r}_3)=V^C({\bf r}_1,{\bf
r}_2,{\bf r}_3)+ V^{conf}({\bf r}_1,{\bf r}_2,{\bf
r}_3)+\sum_{k=1}^6\Delta V_k({\bf r}_1,{\bf r}_2,{\bf r}_3),
\end{equation}
\begin{equation}
\Delta V_1=-\sum_{i=1}^3\frac{A}{8m_i^2}\nabla_i^2r_i+
\sum_{i=1}^3\frac{A}{8m_i^2}\left\{r_i\left[{\bf
p}_i^2-\frac{({\bf p}_i{\bf r}_i)^2}{r_i^2}\right]\right\}_W+
\sum_{i,j=1,i<j}^3\frac{A}{2m_im_j}\left\{r_{ij}\left[{\bf
p}_{ij}^2- \frac{({\bf p}_{ij}{\bf
r}_{ij})^2}{r_{ij}^2}\right]\right\}_W,
\end{equation}
\begin{equation}
\Delta
V_2=\sum_{i,j=1,i\ne j}^3\frac{1}{8m_i^2}\nabla_i^2\left(-\frac{2}{3}
\frac{\alpha_s^{ij}}{r_{ij}}\right),
\end{equation}
\begin{equation}
\Delta V_3=-\sum_{i=1}^3\frac{B}{8m_i^2}{\bf
p}_i^2-\sum_{i,j=1;i<j}^3 \frac{B}{2m_im_j}{\bf
p}_{ij}^2-\sum_{i=1}^3\frac{{\bf p}_i^4}{8m_i^3},
\end{equation}
\begin{equation}
\Delta V_4=\sum_{i,j=1;i<j}^3\frac{1}{2m_im_j}\left\{\frac{2}{3}
\frac{\alpha_s^{ij}}{r_{ij}}\left[{\bf p}_i{\bf p}_j+
\frac{({\bf p}_i{\bf r}_{ij})({\bf p}_j{\bf r}_{ij})}{r_{ij}^2}\right]\right\}_W,
\end{equation}
\begin{equation}
\Delta V_5=-\sum_{i,j=1;i<j}^3\frac{3{\alpha_s^{ij}}^2}{8\pi
m_i^2} \nabla_i^2\frac{\ln(\tilde\mu_{ij}r_{ij})}{r_{ij}},
\end{equation}
\begin{equation}
\Delta V_6=\sum_{i,j=1;i<j}^3\frac{3{\alpha_s^{ij}}^2}{2\pi
m_im_j} \left\{{\bf p}_i{\bf
p}_j\frac{\ln(\tilde\mu_{ij}r_{ij})}{r_{ij}} +\frac{({\bf p}_i{\bf
r}_{ij})({\bf p}_j{\bf r}_{ij})}{r_{ij}^2}
\left(\frac{\ln(\tilde\mu_{ij}r_{ij})}{r_{ij}}-\frac{1}{r_{ij}}\right)
\right\}_W,
\end{equation}
where $\{...\}_W$ denotes the Weyl ordering of operators. Accounting that
the hyperradius $R$ is independent of the order of the quark numbering
we can average the potentials (14)-(19) over the functions $\Psi(R)$ to
find their contributions to the baryon mass. The basic relation used for
this aim is the following:
\begin{equation}
\Delta E_k=\int\frac{d{\mathstrut\bm\rho}
d{\mathstrut\bm\lambda}} {\pi^3}\Psi(R)\Delta
V_k({\mathstrut\bm\rho},{\mathstrut\bm\lambda})\Psi(R).
\end{equation}
Let us note, that the volume element $d{\mathstrut\bm\rho}
d{\mathstrut\bm\lambda}$ can be written in terms of hyperradius
$R$ and the angle $\theta$ ($\rho=R\cos\theta$,
$\lambda=R\sin\theta$, $0\leq\theta\leq\pi/2$):
\begin{equation}
d{\mathstrut\bm\rho}d{\mathstrut\bm\lambda}=(4\pi)^2R^5\sin^2\theta\cos^2\theta
dRd\theta.
\end{equation}

Having the solution of Eq.(11) in the numerical form, we have
calculated all corrections from the interaction operators (14)-(19)
numerically. In Table I we present the values of corresponding
spin-independent matrix elements for the three-quark system
$(Q_1Q_2Q_3)$ with the precision $0.001~GeV$. This is not the accuracy
of the mass spectrum calculation because the different theoretical
uncertainties remain large (see the discussion below). A number of
the Breit-like potentials (14)-(19) is dependent on the relative
quark distances ${\bf r}_{ij}$ which could be expressed directly
through the variables ${\mathstrut\bm\rho}_{ij}$: ${\bf
r}_{ij}={\mathstrut\bm\rho}_{ij}/\alpha_{ij}$. The confinement
contributions and the pure relativistic corrections
$\sum_{i=1}^3{\bf p}_i^4/8m_i^3$ in the potential (13) are
determined in terms of the variables ${\mathstrut\bm\lambda}_{ij}$:
${\bf r}_k=-\gamma_{ij}{\mathstrut\bm\lambda}_{ij}$. The
perturbative and nonperturbative relativistic corrections are
grouped together for the convenience in the expression $\Delta V_3$.
The sign of the first term in the potential (14) differs from the
corresponding expression in Ref.\cite{Brambilla}. The reason
consists in the values of the universal Pauli interaction constant
$\kappa=-1$ and the mixing coefficient $\epsilon=-1$ in the
relativistic quark model fixed from the analysis of heavy quarkonium
masses and radiative decays.

\begin{table}
\caption{\label{t1} Contributions of the spin-independent terms of
the potential (13) to the mass spectrum of the triply and doubly heavy
baryons (in GeV).}
\bigskip
\begin{ruledtabular}
\begin{tabular}{|c|c|c|c|c|c|c|c|c|c|}   \hline
Quark &$<\Delta V_1>$&$<\Delta V_2>$&$<\Delta V_3>$&$<\Delta V_4>$&$<\Delta V_5>$&
$<\Delta V_6>$& Summary   \\
content  &        &      &       &       &       &      & contribution  \\  \hline
$(ccc)$  &0.032   &0.010 &-0.080 &-0.019 &-0.001 &0.001 & -0.057 \\   \hline
$(ccb)$  &0.016   &0.009 &-0.059 &-0.012 &-0.002 &0.0001 & -0.048 \\   \hline
$(bbc)$  &0.007   &0.007 &-0.033 &-0.008 &-0.002 &-0.0003 & -0.029 \\   \hline
$(bbb)$  &0.005   &0.003 &-0.018 &-0.006 &-0.001 &-0.001 & -0.018 \\   \hline
$(ccq)$  &0.037   &0.044 &-0.243 &-0.037 &0.020 &0.013 & -0.166 \\   \hline
$(ccs)$  &0.046   &0.025 &-0.167 &-0.029 &0.004 &0.006 & -0.115 \\   \hline
$(bbq)$  &-0.035   &0.039 &-0.131 &-0.012 &0.013 &0.003 &-0.123 \\   \hline
$(bbs)$  &-0.009  &0.022 &-0.083 &-0.010 &0.001 &0.001 &-0.078 \\   \hline
$(bcq)$  & -0.002  &0.042 &-0.214 &-0.023 &0.016 &0.008 &-0.173 \\   \hline
$(bcs)$  & 0.015  &0.024 &-0.143 &-0.018 &0.002 &0.003 &-0.117\\   \hline
\end{tabular}
\end{ruledtabular}
\end{table}

Another part of the total Hamiltonian depends on the quark spins.
Neglecting the quark orbital momentum we present the spin-dependent part
of the potential
in the simple form \cite{rqm5}:
\begin{equation}
V^{SD}({\bf r}_1,{\bf r}_2,{\bf r}_3)=\sum_{i<j}c_{ij}{\bf S}_i{\bf S}_j
\delta(|{\bf r}_{ij}|),
\end{equation}
\begin{equation}
c_{ij}=\frac{16\pi\alpha_s^{ij}}{9m_im_j}\left[1+\frac{\alpha_s^{ij}}{\pi}
\left(\frac{5}{3}\beta_0-\frac{11}{3}-\left(\frac{m_i-m_j}{m_i+m_j}+
\frac{1}{8}\frac{m_i+m_j}{m_i-m_j}\right)\ln\frac{m_j}{m_i}\right)\right].
\end{equation}
We assume  that the color-magnetic interaction (22) can also be treated
perturbatively as potentials (14)-(19). The matrix element of the
operator (22) can be transformed as follows:
\begin{equation}
\Delta E^{SD}=\frac{64}{9}\sum_{i<j}\frac{\alpha_s^{ij}<{\bf
S}_i{\bf S}_j> \alpha_{ij}^3}{m_im_j\pi}\int_0^\infty\lambda^2
d\lambda |\Psi(\lambda)|^2\times
\end{equation}
\begin{displaymath}
\times\left[1+\frac{\alpha_s^{ij}}{\pi}
\left(\frac{5}{3}\beta_0-\frac{11}{3}-\left(\frac{m_i-m_j}{m_i+m_j}+
\frac{1}{8}\frac{m_i+m_j}{m_i-m_j}\right)\ln\frac{m_j}{m_i}\right)\right].
\end{displaymath}

\begin{table}
\caption{\label{t2} Mass spectrum of ground states of triply heavy
baryons (in GeV). $\{Q_1Q_2\}$ denotes the axial vector two-quark
combination (diquark). The hyperfine splitting has been neglected in
Refs.\cite{kuti,bjorken,jia}.}
\bigskip
\begin{ruledtabular}
\begin{tabular}{|c|c|c|c|c|c|c|}   \hline
Baryon  & Quark content & $J^P$  & This work & \cite{kuti} & \cite{bjorken}
&\cite{jia}   \\  \hline
$\Omega_{ccc}$ & $(ccc)$ & $3/2^+$  & 4.803    &4.79  &4.925 & 4.76  \\  \hline
$\Omega_{ccb}$ & $\{cc\}b$ & $1/2^+$  & 8.018    &---  &--- & ---  \\  \hline
$\Omega^\ast_{ccb}$ & $\{cc\}b$ & $3/2^+$  & 8.025    &8.03  &8.200 & 7.98  \\  \hline
$\Omega_{bbc}$ & $\{bb\}c$ & $1/2^+$  & 11.280    &---  &--- & ---  \\  \hline
$\Omega^\ast_{bbc}$ & $\{bb\}c$ & $3/2^+$  & 11.287    &11.20  &11.480 & 11.19  \\  \hline
$\Omega_{bbb}$ & $(bbb)$ & $3/2^+$  & 14.569    &14.30  &14.760 & 14.37  \\  \hline
\end{tabular}
\end{ruledtabular}
\end{table}

\begin{table}
\caption{\label{t2} Mass spectrum of ground states of doubly heavy
baryons (in GeV). $\{Q_1Q_2\}$ denotes the axial vector two-quark combination
(diquark), $[Q_1Q_2]$ denotes the scalar one.}
\bigskip
\begin{ruledtabular}
\begin{tabular}{|c|c|c|c|c|c|c|c|c|}   \hline
Baryon  & Quark content & $J^P$  & This work & \cite{rqm2}
&\cite{GKLO} & \cite{Roncaglia} & \cite{KKP} &\cite{NT3}  \\  \hline
$\Xi_{cc}$ & $\{cc\}q$ & $1/2^+$  & 3.510    &3.620  &3.478 & 3.66& 3.61& 3.69 \\  \hline
$\Xi^\ast_{cc}$ & $\{cc\}q$ & $3/2^+$  &3.548  &3.727  & 3.61 & 3.74& 3.68&  \\  \hline
$\Omega_{cc}$ & $\{cc\}s$ & $1/2^+$  &3.719 &3.778  & 3.59 &3.74& 3.71& 3.86 \\  \hline
$\Omega^\ast_{cc}$ & $\{cc\}s$ &$3/2^+$  &3.746  &3.872  & 3.69 & 3.82& 3.76&  \\  \hline
$\Xi_{bb}$ & $\{bb\}q$ & $1/2^+$  &10.130 &10.202  & 10.093 & 10.34&  & 10.16 \\  \hline
$\Xi^\ast_{bb}$ & $\{bb\}q$ & $3/2^+$&10.144 &10.237  & 10.133 &10.37&   &  \\  \hline
$\Omega_{bb}$ &$\{bb\}s$ & $1/2^+$  &10.422 &10.359  & 10.18 & 10.37& & 10.34\\  \hline
$\Omega^\ast_{bb}$ & $\{bb\}s$ & $3/2^+$&10.432&10.389  &10.20 &10.40& &  \\  \hline
$\Xi_{cb}$ & $\{cb\}q$& $1/2^+$  & 6.792 &6.933  & 6.82 & 7.04& & 6.96 \\   \hline
$\Xi'_{cb}$ & $[cb]q$ & $1/2^+$  &  6.825 &6.963  & 6.85& 6.99& &  \\  \hline
$\Xi^\ast_{cb}$ & $\{cb\}q$ & $3/2^+$&6.827 &6.980  & 6.90 & 7.06& &  \\  \hline
$\Omega_{cb}$ &$\{cb\}s$ & $1/2^+$  & 6.999 &7.088  & 6.91 & 7.09& & 7.13 \\ \hline
$\Omega'_{cb}$ & $[cb]s$ & $1/2^+$  &7.022 &7.116 &6.93 & 7.06& &  \\  \hline
$\Omega^\ast_{cb}$ & $\{cb\}s$ &$3/2^+$  &7.024&7.130  & 6.99 & 7.12& &  \\  \hline
\end{tabular}
\end{ruledtabular}
\end{table}

To obtain the contribution of the spin-dependent interaction (22)
to the energy spectrum we have to perform the addition of the
quark spins ${\bf S}_i$ to the baryon spin ${\bf S}$. Designating
the heavy quark spins ${\bf S}_1$ and ${\bf S}_2$, we introduce
the spin of a doubly heavy diquark ${\bf S}_{12}= {\bf S}_1+{\bf
S}_2$. Then, the baryon spin is ${\bf S}={\bf S}_{12}+{\bf S}_3$.
Considering that the baryon wave function $\Psi_{S_{12}SS_z}$ is
the eigenfuction for the operators ${\bf S}_{12}^2$, ${\bf S}^2$,
$S_z$, we can express the scalar product $({\bf S}_1{\bf S}_2)$
through the diquark spin ${\bf S}_{12}$:
\begin{equation}
{\bf S}_1{\bf S}_2=\frac{1}{2}\left[S_{12}(S_{12}+1)-\frac{3}{2}\right],
\end{equation}
where $S_{12}=0$ (the scalar diquark) or $S_{12}=1$ (the vector diquark).
For averaging the two other spin dependent terms $({\bf S}_1{\bf S}_3)$ and
$({\bf S}_2{\bf S}_3)$ there is a need to use the unitary transformation
from the wave functions $\Psi_{S_{12}SS_z}$ to the eigenfunctions
$\Psi_{S_{13}SS_z}$ (or $\Psi_{S_{23}SS_z}$) \cite{LL}:
\begin{equation}
\Psi_{S_{12}SS_z}=\sum_{S_{13}}(-1)^{S_1+S_2+S_3+S}
\sqrt{(2S_{12}+1)(2S_{13}+1)}
\Biggl\{{{S_3~S_1~S_{13}}\atop{S_2~S~S_{12}}}\Biggr\}\Psi_{S_{13}SS_z}.
\end{equation}
Corresponding values of $6j$ - symbols are taken from
Ref.\cite{LL}. The presence of operators $({\bf S}_1{\bf S}_3)$
and $({\bf S}_2{\bf S}_3)$ leads to a mixing between states with
different spin $S_{12}$ of the doubly heavy diquark and definite
value $S=1/2$. In the case of the $(bcq)$ and $(bcs)$ baryons we have the
following mixing matrices:
\begin{equation}
{{-0.005~~~~~0.015}\choose{~~0.015~~~-0.020}}[GeV], ~~~
{{-0.004~~~~~0.010}\choose{~~0.010~~~-0.015}}[GeV].
\end{equation}

After the matrix diagonalization we obtain the numerical values for
the masses of the baryons $(bcq)$ and $(bcs)$ $J^P=1/2^+,3/2^+$
which are presented in Table III. In the case of triply and doubly
heavy baryons with two identical heavy quarks $c$ or $b$, the
diquarks $(cc)$ and $(bb)$ have the spin 1 (the axial vector
diquark). The hyperfine mass splittings for doubly heavy $(cc)$ and
$(bb)$ baryons are the following: $\Delta M(\Xi_{cc})=0.038$ GeV,
$\Delta M(\Omega_{cc})=0.027$ GeV, $\Delta M(\Xi_{bb})=0.014$ GeV,
$\Delta M(\Omega_{bb})=0.010$ GeV. They are more than two times
smaller as compared with the splittings obtained in Ref.\cite{rqm4}.
Whereas the masses of triply heavy baryons are in the agreement with
the earlier performed calculations in Refs.\cite{kuti,bjorken,jia}
the obtained masses of doubly heavy baryons lie for the most part
lower than our predictions from \cite{rqm4}. We expect that higher
order corrections both in $\alpha_s$ and $1/m$ could change these
results.

Special attention has to be given to the accuracy of the performed
calculation. In the case of triply heavy baryons the expansion over
$|{\bf p}_{b,c}|/m_{b,c}$ is well defined. Numerical estimate of
relativistic effects gives the following expectation values:
$<{\bf p}_c^2/m_c^2>\approx 0.34$, $<{\bf p}_b^2/m_b^2>\approx 0.09$.
So, the next to leading order contribution ${\bf p}^4/m^4$ leads to
the theoretical uncertainty $\pm 0.030~GeV$ for the charm baryons
and $\pm 0.005~GeV$ for the bottom baryons (see Table II).
The presence of the light quark in the baryon leads
to the increase of the contribution of the relativistic effects
connected with the motion of quarks $q$ and $s$ to the mass spectrum.
Indeed, we can estimate the average value of the square momentum of
light quark using the obtained baryon wave function (see Fig.2).
Introducing the parameter $<{\bf p}_3^2/m_3^2>$ where the brackets
designate the matrix element with the functions of Eq.(9)
we find that in the case of light quark $q$ it has
the value of order unity. So, summary perturbative and nonperturbative
relativistic contribution is sufficiently large (see third column
of Table I). Nevertheless, we suppose that three terms of the
expansion $\epsilon_3= \sqrt{{\bf p}_3^2+m_3^2}\approx m_3+{\bf
p}_3^2/2m_3-{\bf p}_3^4/8m_3^3$ represent the light quark
relativistic energy $\epsilon_3$ with sufficiently high accuracy
(nearly $10\%$). The most significant error is connected with the
relativistic corrections in the second order perturbation theory (PT).
The reduced Green's function $G({\bf R},{\bf R'})$ for Eq.(11)
can be constructed numerically using the solutions of the
Schr\"odinger equation (11) for different principal quantum numbers $n$.
Saturating the sum over $n$ in the expression of the Green's function
by 10-15 excited states, we obtain approximate value $G({\bf R},{\bf R'})$
which can be employed in the second order PT.
Our preliminary estimate of the relativistic contributions in the
second order PT shows that these corrections can comprise near $30\%$
of the contribution $<\Delta V_3>$ (see Table I).

Let us note that exists another possibility to rationalize the relativistic
energy of the light quark. It demands the introduction of the effective
energy (the momentum squared) of the light quark in the bound state.
Indeed, we can use the following transformation of the light quark
energy: $\epsilon_3=\frac{{\bf p}^2_3+m_3^2}{\epsilon_3}\to
\frac{{\bf p}^2_3+m_3^2}{E_3}$ = $\frac{{\bf p}_3^2}{2\tilde m_3}+\frac{m_3^2}
{E_3}$, where $\tilde m_3=E_3/2$, and the light quark effective energy $E_3$ can
be expressed in terms of the baryon mass: $E_3=M_B-m_1-m_2$. In this case
the quantity $\tilde m_3$ plays the role of new effective mass of the
light quark in the baryon and the addendum $\tilde m_3^2/E_3$ is
equivalent to the term $m_3/2$ used in Refs.\cite{NT1,NT2,NT3} if it is
granted that the effective mass $\tilde m_3$ coincides with the
composite quark mass $m_3$. Numerical
value of the light quark effective energy $E_3\approx 0.4\div 0.5~GeV$,
and the shift $m_3^2/E_3\approx 0.21\div 0.22~GeV$ is in the agreement with the
value of the term $m_3/2$ in Refs.\cite{NT1,NT2,NT3}. The appearance of the
effective quantities in this approach evidently leads to the definite theoretical
uncertainty in the baryon mass calculation. So, the presence of the light
quark in the doubly heavy baryon essentially complicates the mass spectrum
calculation with the sufficiently high accuracy. To improve the
theoretical results of Table III the corrections of the second order PT
should be calculated together with the matrix elements of the potential
up to the $1/m^4$ order. The ground state wave function of the triply
charm or bottom baryons is evidently the angle independent. The used
hyperradial approximation is valid for it with high accuracy. For triply
heavy baryons $(ccb)$, $(bbc)$ or for doubly heavy baryons $(Q_1Q_2q)$
this hyperradial approximation is less applicable. So, the dependence of the
baryon wave function on the angular variables in the six-dimensional space
also has to be taken into account.

\acknowledgments The author is grateful to D.Ebert, R.N.Faustov,
V.O.Galkin for discussions and fruitful remarks. The part of this
work was carried out during the visit of the author to the Institute
of Physics of the Humboldt University in Berlin. I am grateful to
M.M\"uller-Preussker and the colleagues from the Institute of
Physics for warm hospitality.
The work is performed under the financial support of the {\it
Deutsche Forschungsgemeinschaft} under contract Eb 139/2-4.

\end{document}